%% file: main.tex
\documentclass[sigconf,nonacm]{acmart}
\setcitestyle{sort}

\renewcommand\footnotetextcopyrightpermission[1]{} 
\usepackage{caption}
\acmConference[DAC'63]{ACM Conference}{July 2026}{Long Beach, CA, USA}

\usepackage{multicol}
\usepackage{multirow}
\usepackage{subfigure}
\usepackage{amsfonts}
\usepackage{amssymb}
\usepackage{textcomp}
\usepackage{xcolor}
\usepackage{enumitem}
\setlist[itemize]{leftmargin=*}
\setlist{
  topsep=0pt,      
  parsep=0pt,       
  itemsep=0pt,      
  partopsep=0pt    
}
\usepackage{hyperref}
\usepackage{bm}
\usepackage[linesnumbered,vlined,ruled]{algorithm2e}
\usepackage{subcaption}
\usepackage{siunitx}
\usepackage{booktabs}
\usepackage{mathtools}
\usepackage{pifont}
\usepackage{tikz}
\usetikzlibrary{quantikz2}
\usepackage{nicematrix}

\SetKw{KwInput}{Input}
\SetKw{KwOutput}{Output}

\newcommand{\nc}{\newcommand}
\nc{\n}{\operatorname}

\nc{\Cal}[1]{\mathcal{#1}} \nc{\fr}[1]{\mathfrak{#1}}
\nc{\Ac}{\Cal{A}} \nc{\Bc}{\Cal{B}} \nc{\Cc}{\Cal{C}} \nc{\Dc}{\Cal{D}} \nc{\Ec}{\Cal{E}}
\nc{\Fc}{\Cal{F}} \nc{\Gc}{\Cal{G}} \nc{\Hc}{\Cal{H}} \nc{\Ic}{\Cal{I}} \nc{\Jc}{\Cal{J}}
\nc{\Kc}{\Cal{K}} \nc{\Lc}{\Cal{L}} \nc{\Nc}{\Cal{N}} \nc{\Oc}{\Cal{O}}
\nc{\Pc}{\Cal{P}} \nc{\Qc}{\Cal{Q}} \nc{\Rc}{\Cal{R}} \nc{\Sc}{\Cal{S}} \nc{\Tc}{\Cal{T}}
\nc{\Uc}{\Cal{U}} \nc{\Vc}{\Cal{V}} \nc{\Wc}{\Cal{W}} \nc{\Xc}{\Cal{X}} \nc{\Yc}{\Cal{Y}}
\nc{\Zc}{\Cal{Z}} 
\nc{\Tr}{\n{Tr}} \nc{\tr}{\n{tr}}
 
\newcommand{\ot}{\otimes}  
\nc{\tht}{\theta} \nc{\om}{\omega}
\nc{\dd}{\,\textnormal{d}}

\nc{\brak}[1]{\langle{#1}\rangle}			\nc{\ketb}[2]{|{#1}\rangle\!\langle{#2}|} 
\nc{\too}{\!\!\to\!\!} 
\nc{\im}{\n{im}}
\newcommand{\CKVQA}{{CutVQA}}

\begin{document}

\title{CutVQA: Co-Designing Circuit Cutting and Architecture Search for Scaling Variational Quantum Algorithms}

\author{Jun Wu}
\email{jun\_wu@mail.ustc.edu.cn}
\orcid{0000-0002-2072-7106}
\affiliation{%
  \institution{School of Computer Science and Technology, University of Science and Technology of China}
  \city{Hefei}
  \state{Anhui}
  \country{China}
}
\author{Jicun Li}
\email{lijicun@mail.ustc.edu.cn}
\affiliation{%
  \institution{School of Computer Science and Technology, University of Science and Technology of China}
  \city{Hefei}
  \state{Anhui}
  \country{China}
}
\author{Jiaqi Yang}
\email{yangjiaqi@mail.ustc.edu.cn}
\affiliation{%
  \institution{School of Computer Science and Technology, University of Science and Technology of China}
  \city{Hefei}
  \state{Anhui}
  \country{China}
}
\author{Wei Xie}
\authornotemark[1]
\email{xxieww@ustc.edu.cn}
\orcid{0000-0001-7163-2521}
\affiliation{%
  \institution{School of Computer Science and Technology, University of Science and Technology of China}
  \city{Hefei}
  \state{Anhui}
  \country{China}
}

\author{Xiang-Yang Li}
\authornote{Wei Xie and Xiang-Yang Li are corresponding authors.}
\email{xiangyangli@ustc.edu.cn}
\orcid{0000-0002-6070-6625}
\affiliation{
  \institution{School of Computer Science and Technology, University of Science and Technology of China}
  \city{Hefei}
  \state{Anhui}
  \country{China}
}
\affiliation{%
  \institution{Hefei National Laboratory, University of Science and Technology of China}
  \city{Hefei}
  \country{China}}

\renewcommand{\shortauthors}{J. Wu et al.}

\begin{abstract}
Circuit cutting enables large quantum circuits to run on small NISQ devices, but it introduces an exponentially high sampling overhead. 
Here, we present \CKVQA, a co-design framework that integrates circuit cutting with quantum architecture search to scale VQAs.
\CKVQA\ performs cutting-aware architecture search and applies subcircuit-level optimization enabled by parameter locality, reducing both reconstruction and training overhead.
Evaluations on two representative VQAs (QAOA and VQE) show that \CKVQA\ matches baseline accuracy while reducing sampling overhead by 2-3 orders of magnitude and shortening training time by at least $50\%$, demonstrating that co-design is essential for scaling VQA execution.
\end{abstract}




\maketitle
\pagestyle{plain}

\section{Introduction}
\input{chapter/1-introduction}
\section{Background and Motivation}
\input{chapter/2-background}
\label{sec:background}
\section{Framework of \CKVQA}
\label{sec:framework}
\input{chapter/4-method}
\section{Experiment Evaluation}
\label{sec:exp}
\input{chapter/5-evaluation}
\section{Conclusion}
\label{sec:conclu}
\input{chapter/7-conclusion}

\begin{acks}
The research is partially supported by Quantum Science and Technology - National Science and Technology Major Project (QNMP) 2021ZD0302900, National Natural Science Foundation of China (Grant No.~62102388), USTC Kunpeng \& Ascend Center of Excellence and CPS-Yangtze Delta Region Industrial Innovation Center of Quantum and Information Technology-MindSpore Quantum Open Fund. 
\end{acks}

\clearpage
\balance
\bibliographystyle{ACM-Reference-Format}
\bibliography{bib}
\end{document}

%% file: chapter/1-introduction.tex
The rapid advances in quantum hardware \citep{arute2019quantum, zhong2020quantum, wu2021strong} have enabled early demonstrations of variational quantum algorithms (VQAs) ~\citep{cerezo2021variational, shang2023shcridinger} on Noisy Intermediate-Scale Quantum (NISQ) processors~\citep{preskill2018quantum}. However, the scalability of VQAs remains fundamentally constrained by two architectural bottlenecks. First, the limited number of physical qubits on a single quantum processing unit (QPU) prevents executing many practically meaningful quantum circuits directly. Second, the high error rates and limited coherence of contemporary devices limit circuit depth, severely degrading algorithmic performance as problem sizes grow. These constraints make it unlikely for near-term quantum hardware to support large-scale VQA workloads without architectural and algorithmic co-optimization.

Distributed quantum computing (DQC) has recently emerged as a promising architectural direction for scaling quantum workloads by coupling multiple smaller QPUs through classical communication links \citep{chen2025circuit, dou2025larqucut}. Circuit cutting techniques, including gate cutting and wire cutting, enable partitioning a large quantum circuit into smaller subcircuits that can be independently executed on different devices. While prior work has demonstrated the feasibility of DQC-based execution~\citep{peng2020simulating, li2023large, tang2021cutqc, mitarai2021constructing, mitarai2021overhead, lowe2023fast, piveteau2023circuit}, including the real device implementation \citep{carrera2024combining, Chong2023experimental}, its practical adoption is still hindered by two system-level challenges.
(1) \textbf{Exponential reconstruction overhead.} circuit cutting reconstruction incurs a sampling cost scaling as $O(C^2)$ with the decomposition’s spectral norm $C$~\citep{piveteau2020advanced}, This overhead grows exponentially with the number of partition points, quickly overwhelming the execution budget. (2) \textbf{Architecture-unaware circuit structures.} Existing VQA ansätze—such as hardware-efficient or problem-inspired circuits—are designed without considering partition locality or communication-aware cost. As a result, they often induce excessive cutting overhead, high inter-device communication, and deep subcircuits that exceed device error budgets.

Quantum Architecture Search (QAS) has been proposed to automatically design high-performing ansätze for VQAs~\citep{grimsley2019adaptive, zhang2022differentiable, du2022quantum}. While QAS improves algorithmic efficiency and expressibility, existing frameworks treat circuit structure design independently of architectural constraints. Current QAS techniques do not incorporate cutting cost, sampling overhead, device qubit limits, multi-QPU execution models during the search process ~\citep{zhang2022differentiable, Martyniuk2024quantum}. Consequently, QAS-generated circuits may achieve good algorithmic performance but remain impractical for distributed execution, leading to poor scalability on NISQ-era quantum hardware.

These observations reveal a fundamental cross-layer gap: \textit{VQA ansatz design, circuit partitioning, and distributed quantum execution have so far been optimized in isolation, leaving substantial inefficiencies when deployed on multi-QPU systems.}

To address this gap, we propose \CKVQA, a cross-layer co-design framework that integrates architecture-aware ansatz generation, optimal circuit cutting, and subcircuit-level optimization to enable scalable VQA execution on distributed quantum systems. \CKVQA\  introduces three key innovations:

\begin{itemize}
\item \textbf{Cutting-aware quantum architecture search.} We incorporate a sampling-overhead and communication-cost model directly into the QAS heuristic, guiding the search toward ansätze that balance algorithmic expressibility with partition locality and distributed executability.
\item \textbf{Architecture-guided circuit partitioning.} We adopt the SMT-based optimization  to determine partition points \citep{Brandhofer2024optimal} that minimize reconstruction overhead under device capacity and communication constraints.
\item \textbf{Subcircuit-localized optimization.} We exploit the parameter locality induced by circuit cutting to update only the subcircuits associated with each parameter, significantly reducing training latency and quantum resource consumption.
\end{itemize}

Our extensive numerical evaluations on VQE and QAOA benchmarks demonstrate that CutVQA reduces sampling overhead by 2-3 orders of magnitude and shortens total training time by $50\%$ relative to baselines, and maintains competitive solution accuracy compared with QAS baselines.
These results highlight that co-optimizing ansatz structure and distributed quantum execution is essential for scaling VQAs under NISQ hardware constraints and provide a systematic methodology for bridging the gap between quantum algorithm design and emerging distributed quantum architectures.


%% file: chapter/2-background.tex
\subsection{Variational Quantum Algorithms and QAS}
VQAs combine Ansätze with classical optimization and represent a leading paradigm for near-term quantum computing~\citep{cerezo2021variational}. A VQA aims to find parameters $\theta^\star$ that minimize
\begin{equation}
    \mathcal{L}(\theta) = \operatorname{tr}(O U(\theta) \rho U(\theta)^\dagger),    
\end{equation}
where $\rho$, $U(\theta)$, and $O$ denote the input state, ansatz, and observable respectively.

The ansatz structure critically determines VQA performance. Traditional approaches include problem-inspired ansätze, which embed domain knowledge for specific tasks~\citep{farhi2014quantum, peruzzo2014variational}, and hardware-efficient ansätze, which respect device connectivity and coherence~\citep{kandala2017hardware}. 
However, both classes may not balance expressibility with hardware limitations.

QAS~\citep{anagolum2024elivagar} offers an automated alternative, exploring circuit design spaces to identify expressive, trainable PQCs. Modern QAS frameworks~\citep{zhang2022differentiable, du2022quantum} employ evolutionary or gradient-based optimization of gate types, connectivity, and parameterization, providing a scalable approach to VQA design as quantum hardware advances.

\subsection{Circuit cutting}
\begin{figure}[!tbp]
\centering
\subfigure[Gate cutting]{\includegraphics[width=0.35\textwidth]{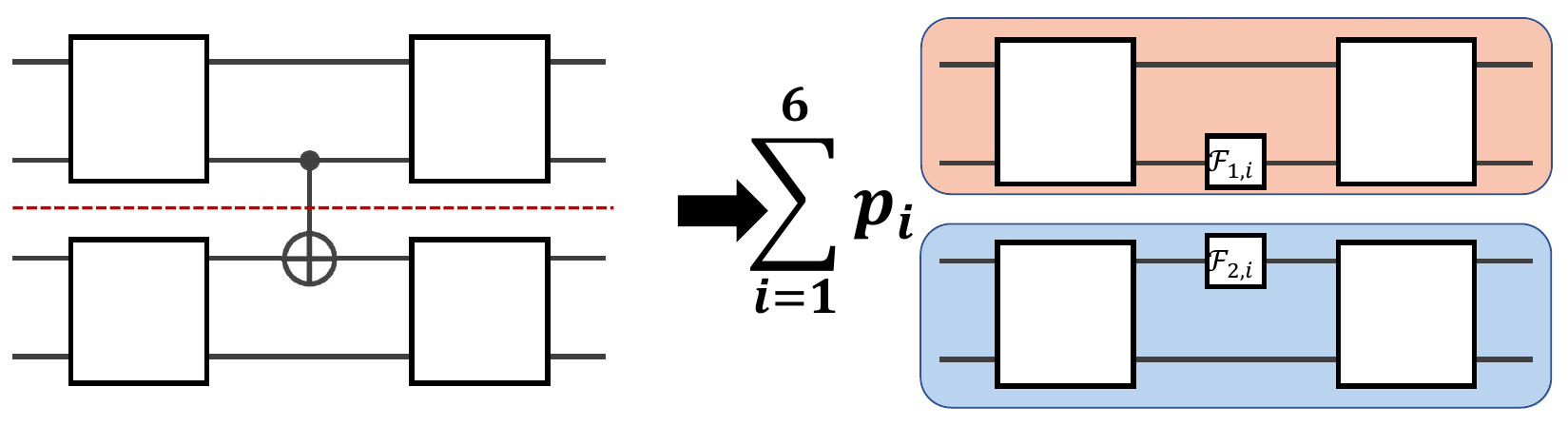}}
\subfigure[Wire cutting]{\includegraphics[width=0.35\textwidth]{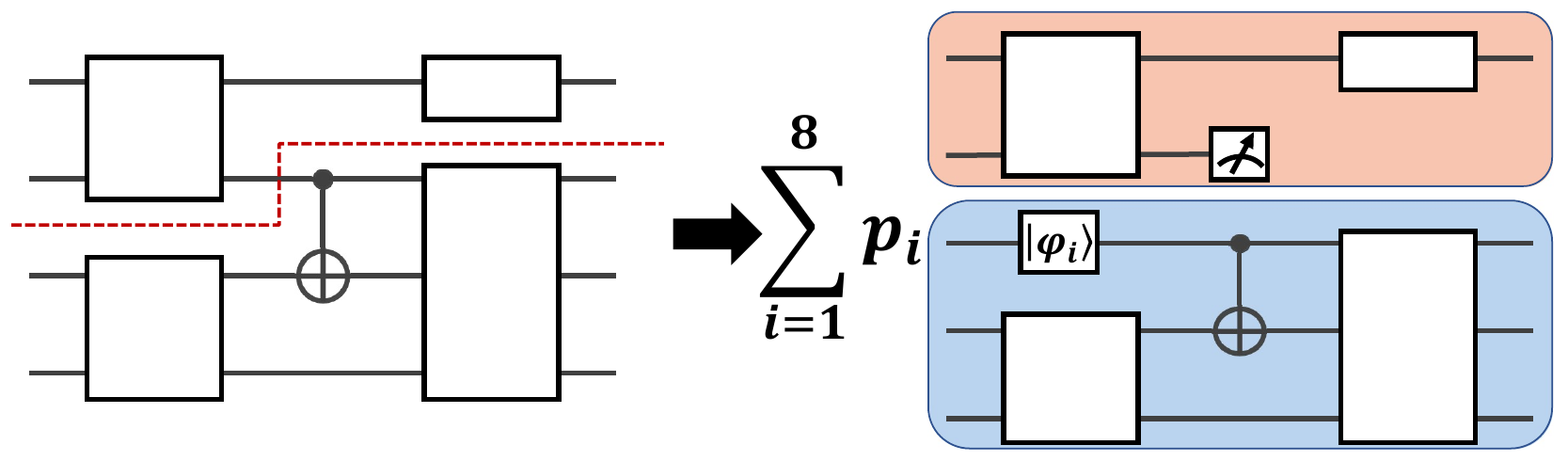}}
\caption{Examples of circuit cutting techniques.}
\label{fig:circuit_cutting}
\Description{}
\end{figure}

Circuit cutting \citep{mitarai2021overhead, piveteau2023circuit} partitions a large quantum circuit into smaller subcircuits that can be executed on limited-qubit devices, followed by classical postprocessing to reconstruct the original computation. 
Two primary approaches are used: \textit{gate cutting} \citep{mitarai2021constructing}, which decomposes two-qubit operations into implementable fragments, and wire cutting, \citep{peng2020simulating, tang2021cutqc, smith2023clifford}, which replaces a qubit line with measurement and state-preparation steps (Figure~\ref{fig:circuit_cutting}). 
A central formalism in circuit cutting is \textit{Quasiprobability Decomposition} (QPD)~\citep{piveteau2022quasiprobability}, which represents a unitary channel $\mathcal{U}$ as
\begin{equation}
\mathcal{U} (\rho) = \sum_i p_i \mathcal{F}_i(\rho),
\end{equation}
for any state $\rho$, where each $\mathcal{F}_i$ is locally implementable and $p_i$ are real coefficients.

A major limitation of circuit cutting is the exponential sampling overhead, which scales as $C^2$ with $C := \sum_i |p_i|$~\citep{piveteau2020advanced}. Intuitively, the cost grows exponentially with the number of partition points. Optimal partitioning strategies have been explored using formal methods, such as SMT-based optimization~\citep{Brandhofer2024optimal}.

\input{chapter/3-motivation}

%% file: chapter/3-motivation.tex
\subsection{Motivation}
VQA scalability is inherently a cross-layer challenge, shaped jointly by the ansatz design at the algorithmic layer and device constraints at the hardware layer. 
The structure of a ansatz directly determines the number of required partitions and the resulting quasiprobability overhead, making ansatz selection a key architectural decision.

\begin{figure}[!tbp]
    \centering
    \includegraphics[width=0.35\textwidth]{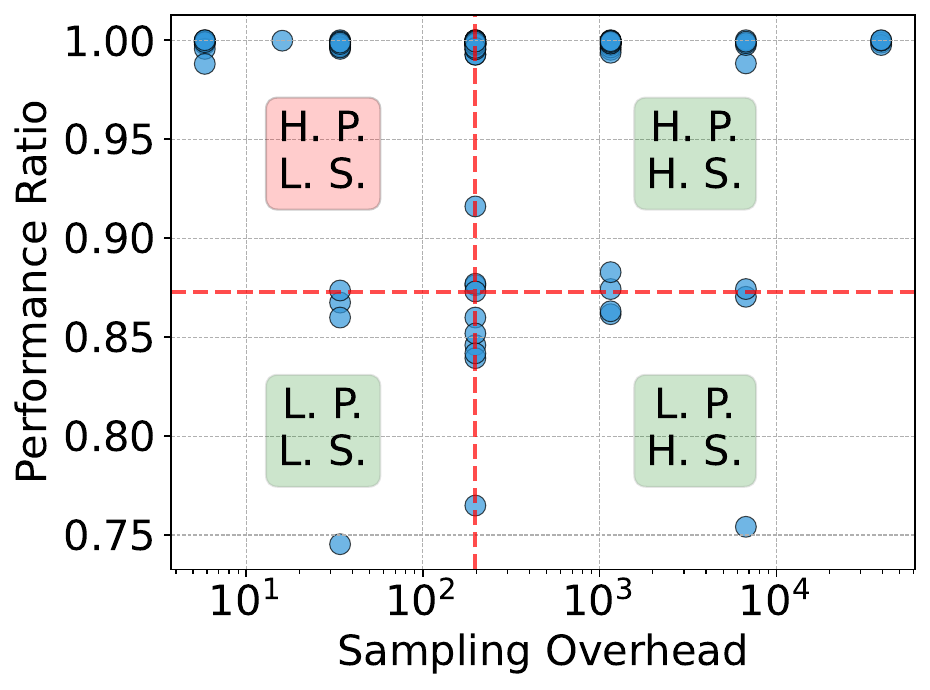}
    \caption{Algorithmic performance versus sampling overhead for various QAOA ansätze on a MaxCut instance. H.P.: High Performance; L.P.: Low Performance; H.S.: High Sampling Overhead; L.S.: Low Sampling.
    }
    \label{fig:motivation-VarousAnsatzPerformance}
    \Description{}
\end{figure}

Figure \ref{fig:motivation-VarousAnsatzPerformance} highlights this gap using QAOA on a 6-node MaxCut instance. 
Different ansätze show large variability in both accuracy and sampling overhead. 
Notably, some designs (red region) lie near a practical Pareto frontier—offering high performance with low reconstruction cost. 
Prior work~\citep{sistla2024towards}, similarly showed that redundant entangling gates substantially inflate reconstruction cost.

These findings motivate a co-design strategy in which ansatz construction explicitly accounts for circuit-cutting overhead, instead of treating partitioning as a post-processing step. 
Furthermore, circuit cutting naturally induces parameter locality, since each subcircuit contains only a subset of variational parameters. Leveraging this structure enables subcircuit-level optimization, allowing the classical optimizer to update only the affected subcircuits and reducing end-to-end VQA training latency.

%% file: chapter/4-method.tex
\begin{figure*}[!tbp]
    \centering
    \includegraphics[width=0.75\textwidth]{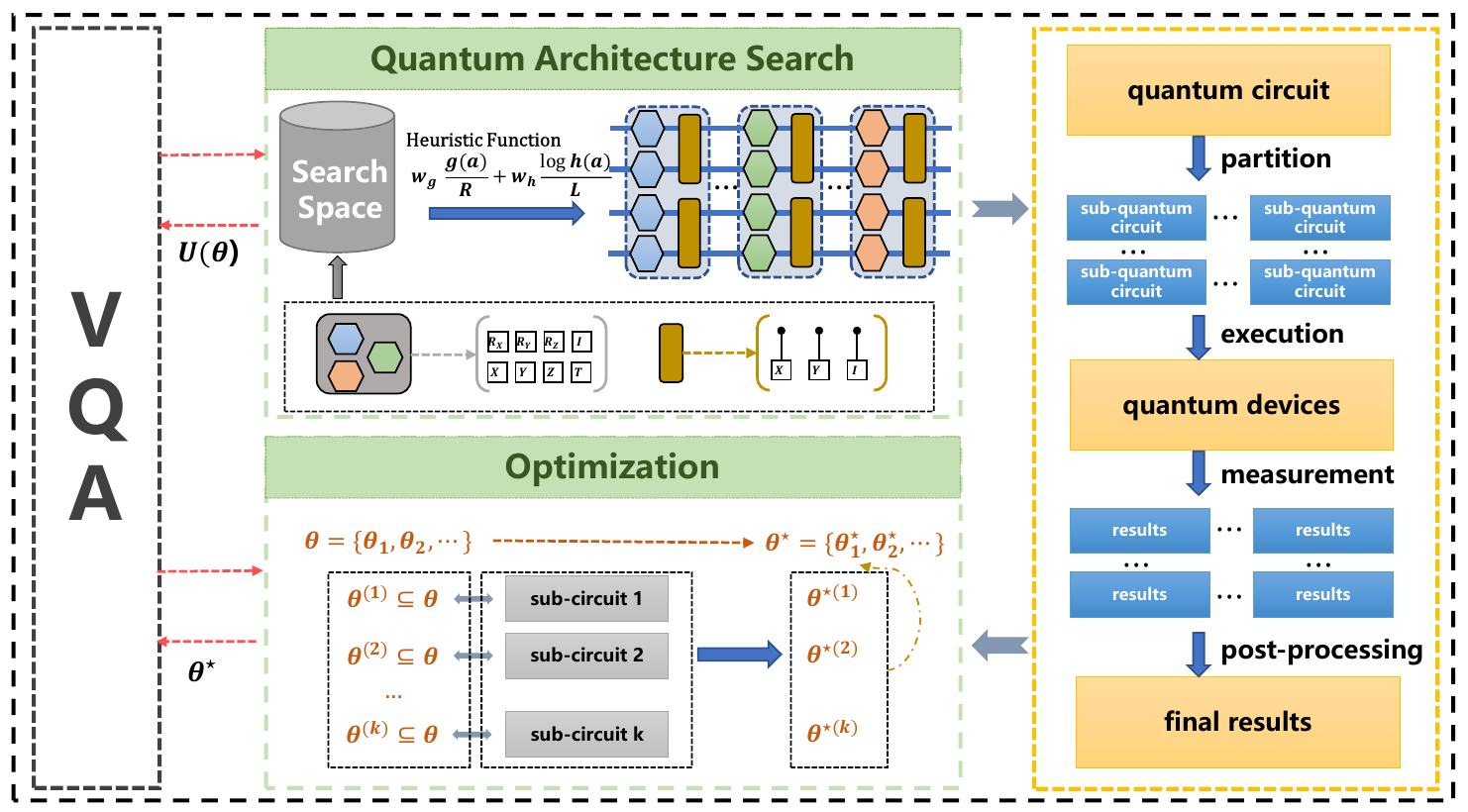}
    \caption{Schematic overview of the \CKVQA\ framework. The quantum architecture search module selects ansätze that balance algorithmic performance with circuit-cutting overhead. The chosen ansatz is partitioned into subcircuits, executed independently, and recombined via classical post-processing. Parameter optimization is localized to individual subcircuits, reducing computational cost and enhancing training efficiency on near-term quantum hardware.}
    \label{fig:CKVQA-framework}
    \Description{}
\end{figure*}

We introduce \CKVQA, a cross-layer co-design framework that unifies quantum architecture search with circuit cutting to enable scalable execution of VQAs. As shown in Figure~\ref{fig:CKVQA-framework}, \CKVQA\ consists of three interdependent modules: (1) QAS for cut-friendly ansätze, (2) Circuit execution via circuit cutting, and (3) parameter-localized optimization for efficient training. 

As shown in Table~\ref{tab:approaches_comparision}, unlike prior work that treats ansatz design, circuit cutting, and VQA training as independent components, CutVQA tightly couples these stages through a distributed execution cost model that informs each optimization step. 

\begin{table}[ht]
\centering
\small
\setlength{\tabcolsep}{2pt}  
\renewcommand{\arraystretch}{0.3} 
\begin{tabular}{lccc}
\toprule
Method & Cutting-Aware & Architecture Search & Local Training \\
\midrule
HEA / UCC & $\times$ & $\times$ & $\times$ \\
QAS (prior) & $\times$& $\checkmark$ & $\times$ \\
Circuit Cutting (prior) & $\checkmark$ & $\times$ & $\times$ \\
\textbf{CutVQA (ours)} & $\checkmark$ & $\checkmark$ & $\checkmark$ \\
\bottomrule
\end{tabular}
\caption{Comparison with prior approaches.}
\label{tab:approaches_comparision}
\end{table}

\subsection{QAS for Cut-Friendly Ansätze}
In \CKVQA, QAS is leveraged to automatically identify ansätze that achieve an optimal balance between these objectives. The procedure, summarized in Algorithm~\ref{alg:qas}, evaluates candidate architectures for sampling overhead and performance, penalizing those exceeding a predefined threshold $\eta$ to favor cut-friendly designs. Iterative selection and population updates converge to an optimal architecture $a^\star$ that balances algorithmic performance and distributed executability.

\begin{algorithm}
    \caption{Genetic Algorithm-based QAS in \CKVQA}
    \label{alg:qas}
    \KwInput{Task specification, single- and two-qubit gate sets $G_{single}, G_{two}$, entanglement pattern $E$, max repetitions $L$, population size $p$, max generations $m$, sampling threshold $\eta$, crossover rate $c_r$, mutation rate $m_r$}\;
    \KwOutput{Optimal quantum architecture $a^\star$}\;

    \textit{Step 1: Generate Search Space} \;
    Construct $\Sc$ using $G_{single}, G_{two}$ and $E$\;

    \textit{Step 2: Initialize Population} \;
    Randomly generate $P = \{a_1, \dots, a_p\}$\;

    \For{$t = 1$ to $m$}{
        \textit{Step 3: Evaluate Fitness} \;
        \For{each $a_i$ in $P$}{
            Compute $h(a_i)$\;
            \eIf{$h(a_i) > \eta$}{$f_i \leftarrow \inf$}{
                Compute $g(a_i)$; $f(a_i) = w_g g(a_i)/R + w_h \log(h(a_i))/L$\;
            }
        }
        
        \textit{Step 4: Evolution} \;
        Select parents $P_{parents}$ from $P$\;
        Generate offspring $P_{offspring}$ via crossover ($c_r$) and mutation ($m_r$)\;
        Form new population $P$ by combining offspring and elites\;
    }

    \textbf{return} $a^\star = \arg\min_i f_i$ in $P$\;
\end{algorithm}

\subsubsection{Search Space Definition} 
The QAS search space $\Sc$ consists of candidate ansatz templates constructed from a predefined gate set and constrained. Specifically, it is defined by: (1) single-qubit gates $G_{\text{single}}$, e.g., $R_{\n{x}}$, $R_{\n{y}}$, $R_{\n{z}}$; (2) two-qubit gates $G_{\text{two}}$, e.g., CX and CZ; and (3) the repetition limit $L$ specifying the maximum ansatz layers. Formally,
\begin{equation}
\Sc = \left\{a_i \mid a_i = \prod_{j=1}^{L} U_j^{(\text{single})}(\theta) U_j^{(\text{two})}\right\},
\end{equation}
where $U_j^{(\text{single})}$ and $U_j^{(\text{two})}$ denote sequences of single- and two-qubit gates, respectively. 
To facilitate efficient storage and manipulation, each architecture is encoded as an integer sequence representing gates and entanglement patterns. For instance, consider a 3-qubit task with $G_{\text{single}} = {R_x, R_y, R_z}$, $G_{\text{two}} = {\text{CX}, \text{CZ}, I}$, and an entanglement pattern [[0,1], [1,2]]. The integer encoding [[0,0,0,0,0],[1,1,1,1,2]] corresponds to the quantum circuit 
\begin{equation}
    \n{CZ}(0,1) \bigotimes_{i=0}^3 R_{\n{y}}(\theta_i) \n{CX}(0,1) \n{CX}(1,2) \bigotimes_{j=0}^2R_{\n{x}}(\theta_j) .
\end{equation}
where each integer maps to a specific gate in the sequence, enabling compact representation and efficient traversal during the architecture search.

The QAS search space $\Sc$ grows combinatorially with the number of qubits $n$, gate sets $G_{\text{single}}$, $G_{\text{two}}$, and depth $L$. For each layer, there are $|G_{\text{single}}|^n \cdot |G_{\text{two}}|^E$ possible arrangements, where $E$ is the number of entangling gates. The total space scales as: $|\Sc| = (|G_{\n{single}}|^n \cdot |G_{\n{two}}|^E)^L$, growing exponentially with $n$ and $L$, which motivates heuristic search methods like genetic algorithms to explore promising architectures efficiently.

\subsubsection{Evaluation Metrics} 
Candidate ansätze are evaluated using metrics aligned with VQA objectives: algorithmic performance $g(a)$ and sampling overhead $h(a)$. We design the following heuristic cost function that combines these two metrics to guide search:
\begin{equation}
\label{eq:heuristic_function}
f(a) = w_g \frac{g(a)}{R} + w_h \frac{\log(h(a))}{L},
\end{equation}
where $R$ normalizes $g(a)$ to $[0,1]$ (e.g., half the number of edges for Max-Cut QAOA or the Hartree–Fock energy for molecular VQE), and $L$ normalizes $\log(h(a))$, typically set to the ansatz repetition limit used to generate the search space.

The logarithm of $h(a)$ compresses its exponential growth with cut points, aligning its scale with normalized performance and ensuring balanced contributions in $f(a)$. A linear combination of normalized performance and overhead offers a simple, interpretable trade-off, tunable via $w_g$ and $w_h = 1-w_g$. Sampling overhead $h(a)$ is computed using SMT-based optimal gate and wire cutting~\citep{Brandhofer2024optimal}.

\subsubsection{Search Strategy} 
\CKVQA\ employs an \textit{evolutionary search}~\citep{gepp2009review, katoch2021review, rasconi2019innovative} to explore the ansatz design space. Each candidate ansatz is treated as an individual in a population, and the search iterates through: (i) selection, favoring high-scoring individuals; (ii) crossover, recombining structures of top candidates; and (iii) mutation, introducing controlled variations to maintain diversity. This balance of exploitation and exploration efficiently navigates the high-dimensional design space, increasing the likelihood of discovering ansätze with both high algorithmic performance and low sampling overhead. 
Evolutionary search is particularly suitable for ansatz design because it approximates global optimization in combinatorial spaces without requiring gradients, which are often unavailable or costly for quantum circuits. Stochastic mutation and crossover ensure broad coverage of the search space, while selection guides convergence toward high-quality, cut-friendly architectures. Prior studies~\citep{gepp2009review, rasconi2019innovative} confirm robust convergence behavior in similar combinatorial optimization tasks. 

To further improve efficiency, \CKVQA\ incorporates two heuristics: (i) Sampling Overhead Threshold, pruning candidates with $h(a)>\eta$ to avoid infeasible designs; and (ii) Random Subset Selection, limiting exploration to a randomly chosen subset of nodes per generation, preserving diversity while controlling complexity. Together, these strategies enable rapid convergence to high-quality, hardware-aware ansätze suitable for NISQ devices.

\subsubsection{Search Scalability Analysis}
In practice, this means that the majority of search time is \textit{classical pre-processing}, analogous to the compilation phase in classical circuit design, and does not occupy costly quantum runtime. Moreover, the search can leverage several practical techniques to further reduce classical computational overhead, such as parallel evaluation of candidate sub-circuits, reusing previously simulated sub-circuit results, and pruning the search space based on heuristic cost functions. As a result, the combined QAS and circuit cutting framework remains tractable for moderate- to large-scale VQA tasks, while maintaining acceptable sampling costs for classical simulation.

\subsection{Circuit Execution via Circuit Cutting}
Circuit cutting enables the execution of quantum circuits beyond the hardware qubit limits by partitioning them into smaller, independently executable subcircuits. We focus on gate cutting, which decomposes two-qubit gates to facilitate circuit partitioning. 
Given an \(n\)-qubit quantum circuit \(U(\theta)\) and a device with a maximum qubit capacity of \(m\), the determination of optimal partition points is formulated as an SMT optimization problem, following the approach in Ref.~\citep{Brandhofer2024optimal}. Solving this problem identifies cutting positions that minimize the sampling overhead introduced by circuit cutting.

After partitioning, each subcircuit is then executed independently on the quantum device. The final output is reconstructed via classical post-processing, combining partial results according to quasiprobability coefficients and the partition structure, thereby preserving an unbiased representation of the original computation.

\subsection{Parameter-Localized Optimization}
\label{sec:sub_opt}


Circuit cutting partitions the original quantum circuit into smaller subcircuits, each containing only a subset of parameters. This enables parameter-localized optimization, where each parameter $\theta_i$ is updated using only the subcircuits that depend on it, avoiding redundant full-circuit evaluations and reducing computational cost while maintaining unbiased gradient estimates.

Exploiting parameter locality, let $\mathcal{S}i$ denote the set of subcircuits containing $\theta_i$, and $C\ell$ the $\ell$-th subcircuit. The gradient of the VQA cost function $\mathcal{L}(\boldsymbol{\theta})$ w.r.t. $\theta_i$ decomposes as
\begin{equation}
\label{eq:gradient_decomposition}
\frac{\partial \mathcal{L}(\boldsymbol{\theta})}{\partial \theta_i}
= \sum_{C_\ell \in \mathcal{S}i} w\ell \frac{\partial \mathcal{L}\ell(\boldsymbol{\theta}\ell)}{\partial \theta_i},
\end{equation}
where $\mathcal{L}\ell$ is the subcircuit contribution and $w\ell$ the reconstruction weight. Circuit cutting ensures this decomposition is unbiased~\citep{piveteau2022quasiprobability}:
\begin{equation}
\mathbb{E}\Bigg[\sum_{C_\ell \in \mathcal{S}i} w\ell \frac{\partial \mathcal{L}_\ell}{\partial \theta_i} \Bigg]
= \frac{\partial \mathcal{L}(\boldsymbol{\theta})}{\partial \theta_i}.
\end{equation}
Thus, standard gradient-based optimizers (e.g., gradient descent, Adam) converge to the same stationary points as full-circuit training, with lower computational overhead.

The optimization procedure consists of two modules:  
\textbf{(i) Parameter Identification Module:} For each $\theta_i$, the relevant subset $\mathcal{S}_i$ of subcircuits is identified. This can be efficiently implemented using a disjoint-set data structure~\citep{galler1964improved} in polynomial time.  
\textbf{(ii) Subcircuit-Based Gradient Update Module:} Gradients with respect to $\theta_i$ are computed solely from subcircuits in $\mathcal{S}_i$, and parameters are updated using standard optimizers.  
If a parameter $\theta_i$ appears in $k$ out of $L$ total subcircuits, the expected acceleration is $O(k/L)$. 

While the Quantum Architecture Search (QAS) procedure can be computationally intensive, its combination with circuit cutting significantly mitigates the dependence on quantum hardware. Specifically, large-scale quantum circuits that are classically intractable can be partitioned into smaller sub-circuits of manageable size, which can be efficiently simulated classically. Consequently, QAS operates predominantly on these classically simulable sub-circuits, reducing the need for direct execution on expensive quantum devices.


%% file: chapter/5-evaluation.tex
\begin{table*}[!htbp]
    \setlength{\abovecaptionskip}{0pt}  
\setlength{\belowcaptionskip}{2pt}  
    \caption{Performance of Different VQAs}
    \label{tab:benchmarksPerformance}
    \centering
    \renewcommand{\arraystretch}{0.3} 
    \begin{tabular}{c c c c c cc cc cc}
        \toprule
        \multirow{2}{*}{Benchmark} &
        \multirow{2}{*}{Type} &
        \multirow{2}{*}{$n$} &
        \multirow{2}{*}{$m$} &
        \multirow{2}{*}{$L$} &
        \multicolumn{2}{c}{\textbf{\CKVQA}} &
        \multicolumn{2}{c}{\textbf{QAS}} &
        \multicolumn{2}{c}{\textbf{HEA}}
        \\
        \cmidrule(lr){6-7}
        \cmidrule(lr){8-9}
        \cmidrule(lr){10-11}
        & & & & &
        Perf. & Over. &
        Perf. & Over. &
        Perf. & Over.
        \\
        \midrule

        \multirow{12}{*}{QAOA}
        & \multirow{4}{*}{REG}
        & 6 & 3 & 5 & $r=0.837$ & 2.6 & 0.839 & 1439.2 & 0.738 & $9^5\approx 6\times 10^4$ \\
        & & 8 & 4 & 5 & $r=0.85$ & 3.4 & 0.89  & 837.0  & 0.722 & $9^5\approx 6\times 10^4$ \\
        & & 10 & 5 & 5 & $r=0.83$ & 3.4 & 0.83  & 259.3  & 0.616 & $9^5\approx 6\times 10^4$ \\
        & & 12 & 6 & 5 & $r=0.78$ & 1.8 & 0.80  & 202.5  & 0.579 & $9^5\approx 6\times 10^4$ \\
        \cmidrule(lr){2-11}

        & \multirow{4}{*}{ERD}
        & 6 & 3 & 5 & 0.862 & 3.4 & 0.872 & 779.4 & 0.775 & $9^5\approx 6\times 10^4$ \\
        & & 8 & 4 & 5 & 0.903 & 2.6 & 0.919 & 901.8 & 0.683 & $9^5\approx 6\times 10^4$ \\
        & & 10 & 5 & 5 & 0.859 & 1.8 & 0.862 & 893.7 & 0.708 & $9^5\approx 6\times 10^4$ \\
        & & 12 & 6 & 5 & 0.844 & 3.4 & 0.824 & 326.5 & 0.648 & $9^5\approx 6\times 10^4$ \\
        \cmidrule(lr){2-11}

        & \multirow{4}{*}{BAR}
        & 6 & 3 & 5 & 0.772 & 3.4 & 0.777 & 981.0 & 0.739 & $9^5\approx 6\times 10^4$ \\
        & & 8 & 4 & 5 & 0.840 & 3.4 & 0.850 & 397.8 & 0.723 & $9^5\approx 6\times 10^4$ \\
        & & 10 & 5 & 5 & 0.842 & 3.4 & 0.836 & 326.5 & 0.681 & $9^5\approx 6\times 10^4$ \\
        & & 12 & 6 & 5 & 0.822 & 2.6 & 0.803 & 183.2 & 0.653 & $9^5\approx 6\times 10^4$ \\
        \midrule

        \multirow{4}{*}{VQE}
        & H$_2$ & 4 & 2 & 5 & $\delta=3.8\times 10^{-9}$ & 3.4 & $5.66\times10^{-9}$ & 461.7 & $2.64\times 10^{-5}$ & $9^5\approx 6\times10^4$ \\
        & LiH  & 6 & 3 & 5 & $\delta=0.006$ & 1.8 & 0.014 & 196.2 & 0.007 & $9^5\approx 6\times10^4$ \\
        & F$_2$ & 8 & 4 & 5 & $\delta=0.011$ & 3.0 & 0.08 & 381.7 & 0.027 & $9^5\approx 6\times10^4$ \\
        & H$_2$O & 8 & 4 & 5 & $\delta=0.0161$ & 1.8 & 0.017 & 381.8 & 0.022 & $9^5\approx 6\times10^4$ \\
        \bottomrule
    \end{tabular}
    \vspace{2mm}
    \begin{minipage}{\textwidth}
        \footnotesize
        \textbf{Note.} 
        $n$: total qubits. 
        $m$: device qubit limit. 
        $k$: ansatz repetition. 
        $r$: QAOA solution ratio from the optimal solution. 
        $\delta$: VQE energy difference from reference.
        Pref.: Algorithmic performance.
        Over.: Sampling overheads     
    \end{minipage}
\end{table*}
\subsection{Experimental Setup}
Experiments ran on AMD EPYC 9654 CPUs (96 cores, 192 threads, 2.4GHz) with 512GB DDR5-4800MHz memory, under Ubuntu 22.04.3 LTS. We utilized the quantum software development kit
Mindquantum~\citep{xu2024mindspore} and Qiskit~\citep{qiskit} in our experiments. The updated code is available at ~\citep{code_mind_CutVQA}.

\CKVQA\ was evaluated on VQE~\citep{peruzzo2014variational} for ground-state energies of $\text{H}_2$, LiH, and $\text{F}_2$, and on QAOA~\citep{farhi2014quantum} for Max-Cut on random graphs from m-Regular (REG, $m=3$)~\citep{steger1999generating}, Erdős–Rényi (ERD, $p=0.3$)~\citep{erdds1959random}, and Barabási–Albert (BAR, $m=3$)~\citep{barabasi1999emergence} models.

Candidate ansätze were built from single- and two-qubit gates $G_{single} = \{ R_{\n{x}} , R_{\n{y}}, R_{\n{z}}\}, G_{two} = \{I\ot I, \n{CZ}, \n{CX}\}$ with linear entanglement patterns allowing two-qubit gates only between neighboring qubits.
All candidates were trained with the same iterations, random seeds, optimizer, and learning rate. Subcircuit sampling during circuit cutting was identical, ensuring unbiased gradients and objective values. Baseline QAS used the same evolutionary strategy but with fitness $f(a) = g(a)$ ($w_g = 1, w_h = 0$), i.e., considering only algorithmic performance.

We measured \textbf{Sampling Overhead}, \textbf{VQE Accuracy} (energy deviation), and \textbf{QAOA Performance} (cut-value ratio). Each experiment was repeated ten times unless stated otherwise to mitigate stochastic effects.

\subsection{The performance of \CKVQA}

\subsubsection{Effectiveness of QAS in \CKVQA}

We assess the QAS module by comparing \CKVQA-discovered ansätze with two standard baselines: hardware-efficient ansätze (HEA) and conventional QAS. Experiments are performed on both VQE and QAOA, with all methods restricted to at most five ansatz blocks to ensure fairness.
Table~\ref{tab:benchmarksPerformance} shows algorithmic accuracy and sampling overhead. Note, these two metric are the average value among ten experiments. \CKVQA\ consistently achieves comparable or better performance than HEA and standard QAS while reducing sampling overhead by 2–3 orders of magnitude. This reduction stems from jointly optimizing for expressiveness and cut-induced sampling cost, enabling \CKVQA\ to avoid structures that trigger exponential reconstruction overhead. As a result, \CKVQA\ delivers NISQ-feasible circuit designs with substantially lower resource requirements.

\subsubsection{Effectiveness of Optimization Strategies in \CKVQA}
\begin{figure}[!tbp]
    \centering
    \subfigure[Optimization time vs. problem size]{
    \label{fig:exp-2-eff-opt-a}
    \includegraphics[width=0.225\textwidth]{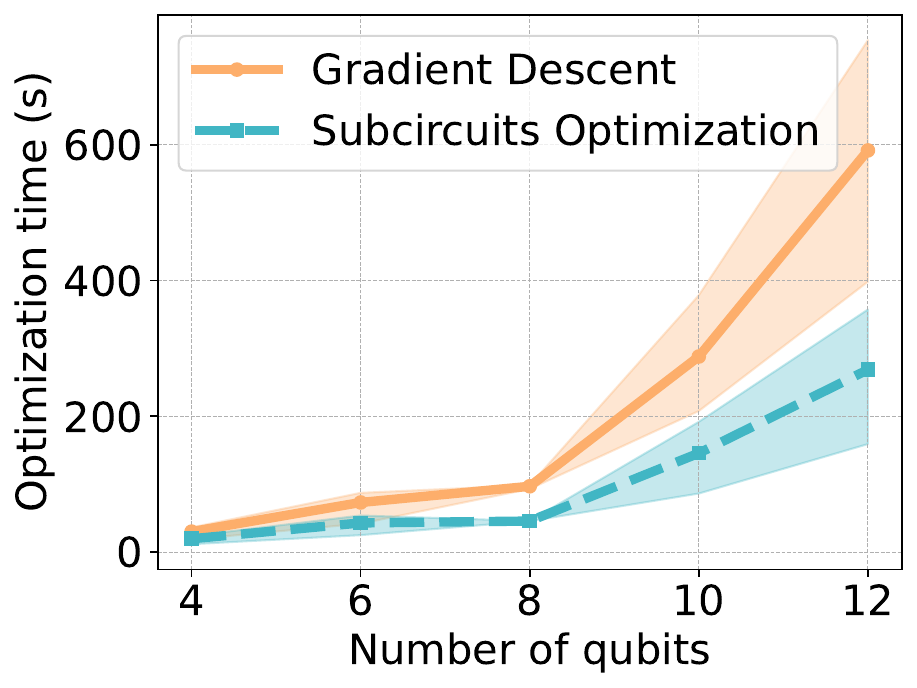}
    }
    \subfigure[Optimization time vs. maximum available qubits]{
    \label{fig:exp-2-eff-opt-b}
    \includegraphics[width=0.225\textwidth]{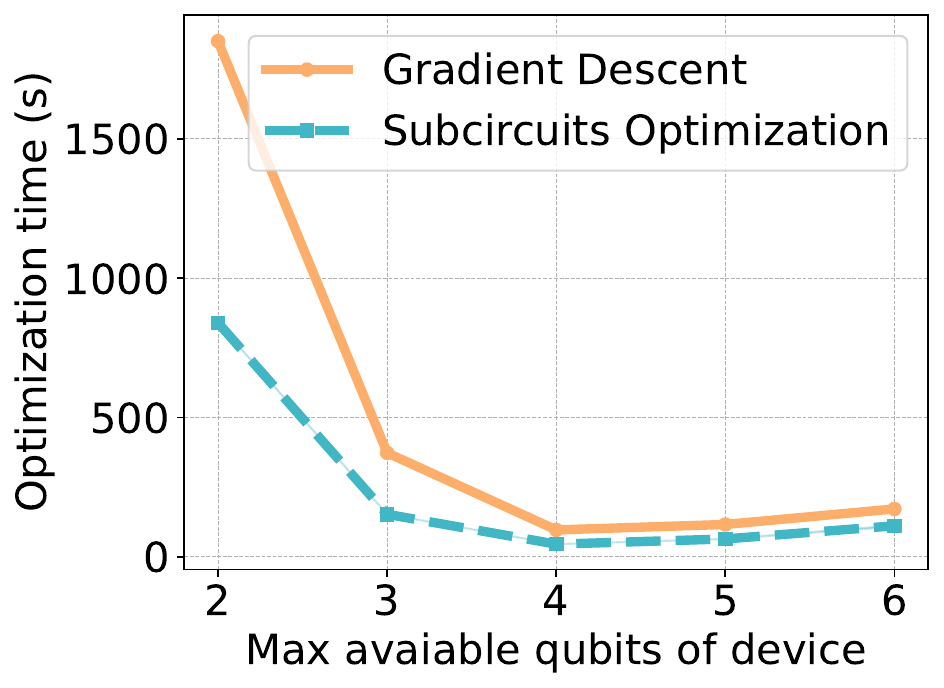}
    }
    \caption{Efficiency of \CKVQA’s parameter-localized optimization. 
    }
    \label{fig:exp-2-eff-opt}
    \Description{}
\end{figure}
We next evaluate the optimization strategy in Sec.~\ref{sec:sub_opt} using QAOA for Max-Cut on 3-regular graphs. The baseline is standard gradient descent, and the primary metric is total optimization time.
Fig.~\ref{fig:exp-2-eff-opt-a} shows that the subcircuit-based optimization achieves increasing speedups as the circuit grows, demonstrating improved scalability for deeper or larger circuits. Fig.~\ref{fig:exp-2-eff-opt-b} fixes the graph size to 8 nodes and varies the qubit budget; the gains are most significant under restricted hardware, highlighting the synergy between QAS and localized optimization in constrained settings. Overall, the results confirm that \CKVQA’s optimization strategy substantially reduces training cost while preserving solution quality.

\subsection{Sensitivity of \CKVQA}
\subsubsection{Number of available qubits}
\begin{figure}[!tbp]
    \centering
    \subfigure[QAOA]{
    \includegraphics[width=0.225\textwidth]{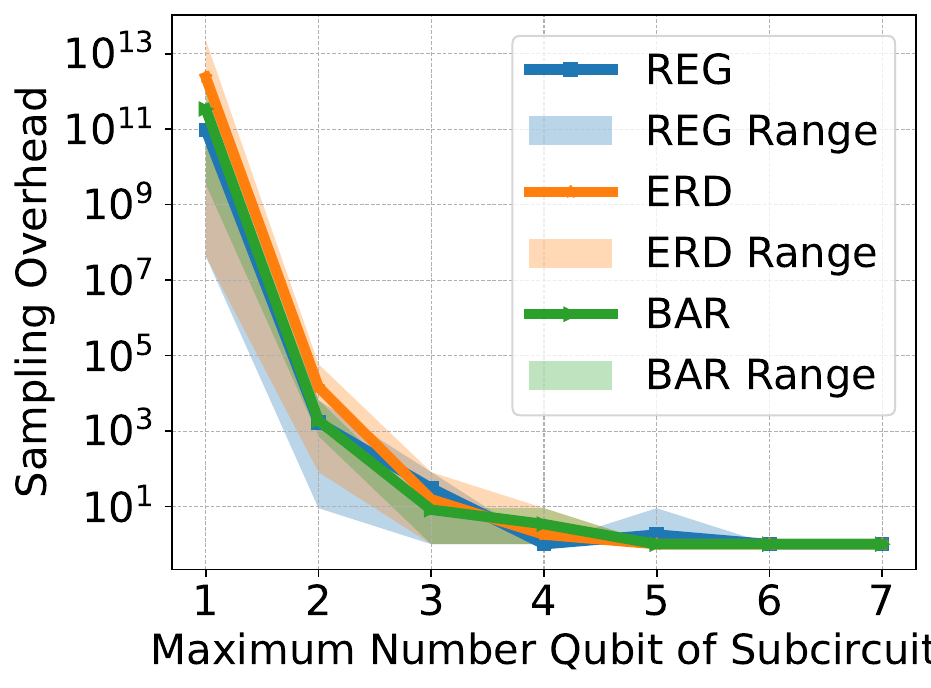}
    }
    \subfigure[VQE]{
    \includegraphics[width=0.225\textwidth]{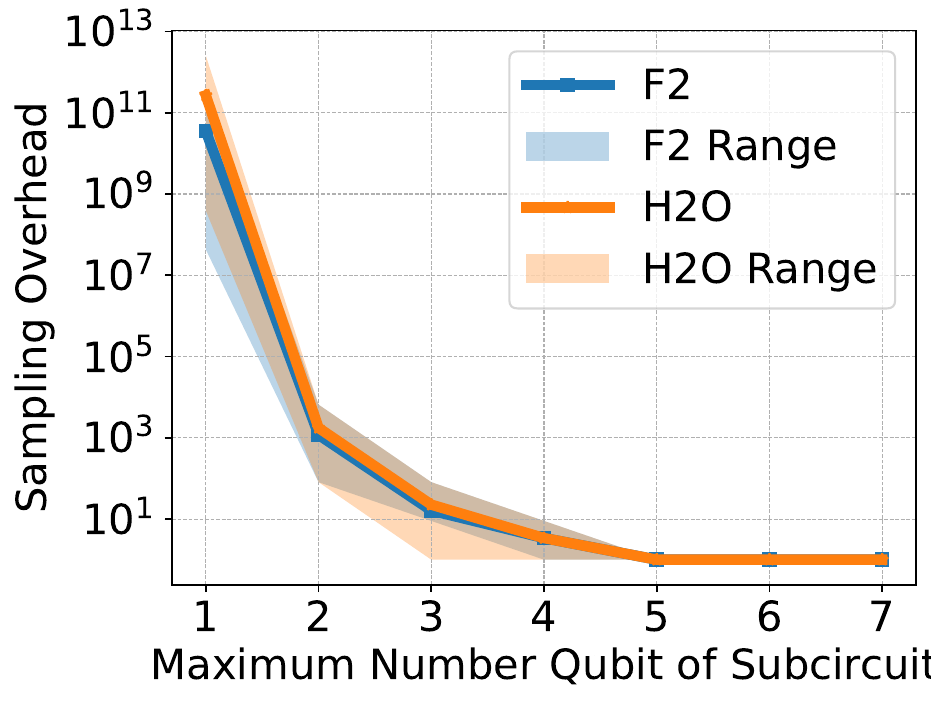}
    }
    \caption{Sampling overhead vs. ansatz depth.}
    \label{fig:exp4-device_configuration}
    \Description{}
\end{figure}

We first examine how device qubit capacity affects circuit-cutting–based VQA execution. Let $m$ be the maximum available qubits, which bounds subcircuit size. Fixing the problem to 8 qubits, we vary $m\in[1,7]$.
Fig.~\ref{fig:exp4-device_configuration} shows that sampling overhead drops rapidly as $m$ increases. When $m \ge 3$, the overhead falls to $\sim10^{1}$, indicating that the partition becomes almost trivial and reconstruction cost is negligible. Thus, even modest device capacities enable efficient cutting, substantially reducing the resource cost of executing larger circuits on NISQ hardware. When each subcircuit fits 3 qubits, sampling overhead becomes negligible; thus mid-size NISQ devices already support efficient cut-based execution.

\subsubsection{Ansatz layer}
\begin{figure}[!tbp]
    \centering
    \subfigure[QAOA]{
    \includegraphics[width=0.225\textwidth]{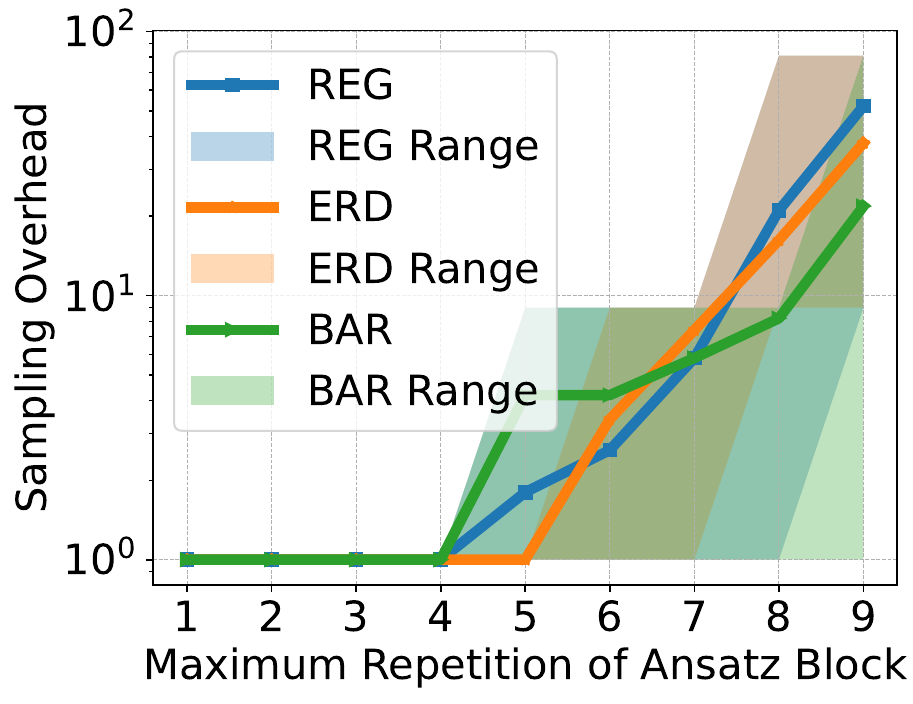}
    }
    \subfigure[VQE]{
    \includegraphics[width=0.225\textwidth]{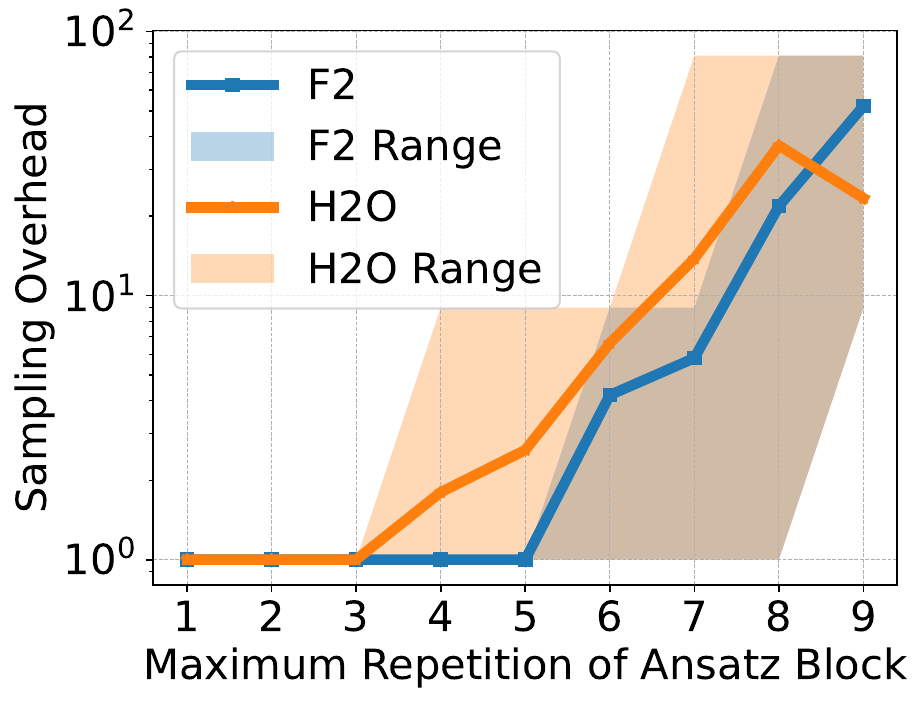}
    }
    \caption{Impact of maximum ansätze repetitions.}
    \label{fig:sensitivity_ansatz_layers}
    \Description{}
\end{figure}

We next study the impact of ansatz depth, which governs expressiveness and computational cost. Fig.~\ref{fig:sensitivity_ansatz_layers} summarizes the results.
Across all configurations, sampling overhead stays low as the number of layers grows. This demonstrates that \CKVQA’s search and partitioning strategy effectively constrains reconstruction cost regardless of circuit depth. As a result, execution time—not sampling—becomes the dominant factor, emphasizing the importance of selecting ansatz depth that balances accuracy and runtime. Since \CKVQA\ keeps reconstruction cost stable across circuit depths, designers may choose deeper ansätze when accuracy is critical—the dominant cost shifts to execution time instead of sampling.

\subsubsection{The coefficients in heuristic function}
\begin{figure}[!tbp]
    \centering
    \subfigure[QAOA]{
    \includegraphics[width=0.225\textwidth]{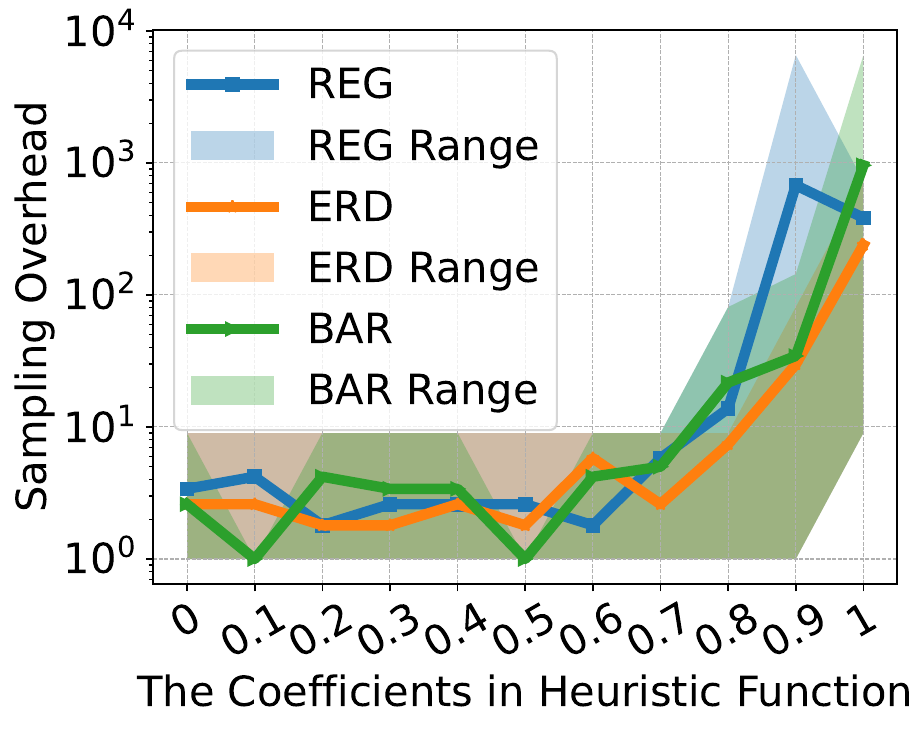}
    }
    \subfigure[VQE]{
    \includegraphics[width=0.225\textwidth]{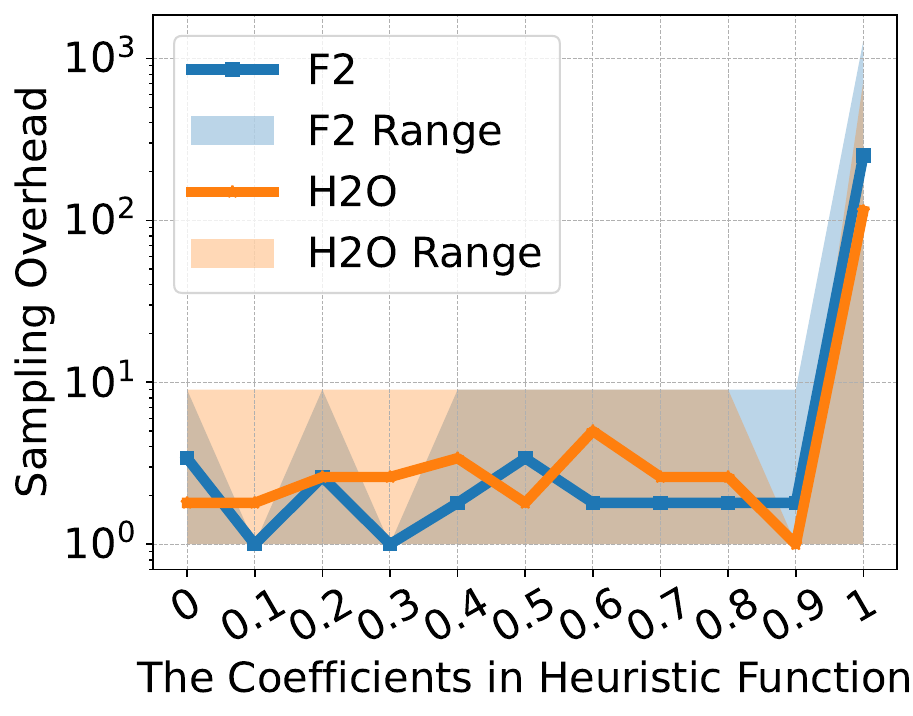}
    }
    \caption{Impact of heuristic function coefficients.}
    \label{fig:exp-4-sensitivity-coefficients}
    \Description{}
\end{figure}

Finally, we evaluate the sensitivity of the heuristic in Eq.~\ref{eq:heuristic_function}. We vary $w_g\in[0, 1]$ and set $w_h=1-w_g$.
As shown in Fig.~\ref{fig:exp-4-sensitivity-coefficients}, sampling overhead increases with $w_g$. Assigning more weight to algorithmic performance leads the search toward deeper or more entangling ansätze, improving solution quality but at a higher reconstruction cost. This behavior confirms that the heuristic provides fine-grained control over the performance–cost trade-off, enabling designers to tune \CKVQA\ based on available hardware budgets and application requirements. The coefficients $(w_g, w_h)$ offer direct control of the accuracy–cost trade-off: higher $w_g$ prioritizes solution quality, while higher $w_h$ enforces cut-friendly architectures suited to tight hardware budgets.

\subsection{Noisy simulation}
\begin{figure}[!tbp]
    \centering
    \includegraphics[width=0.40\textwidth]{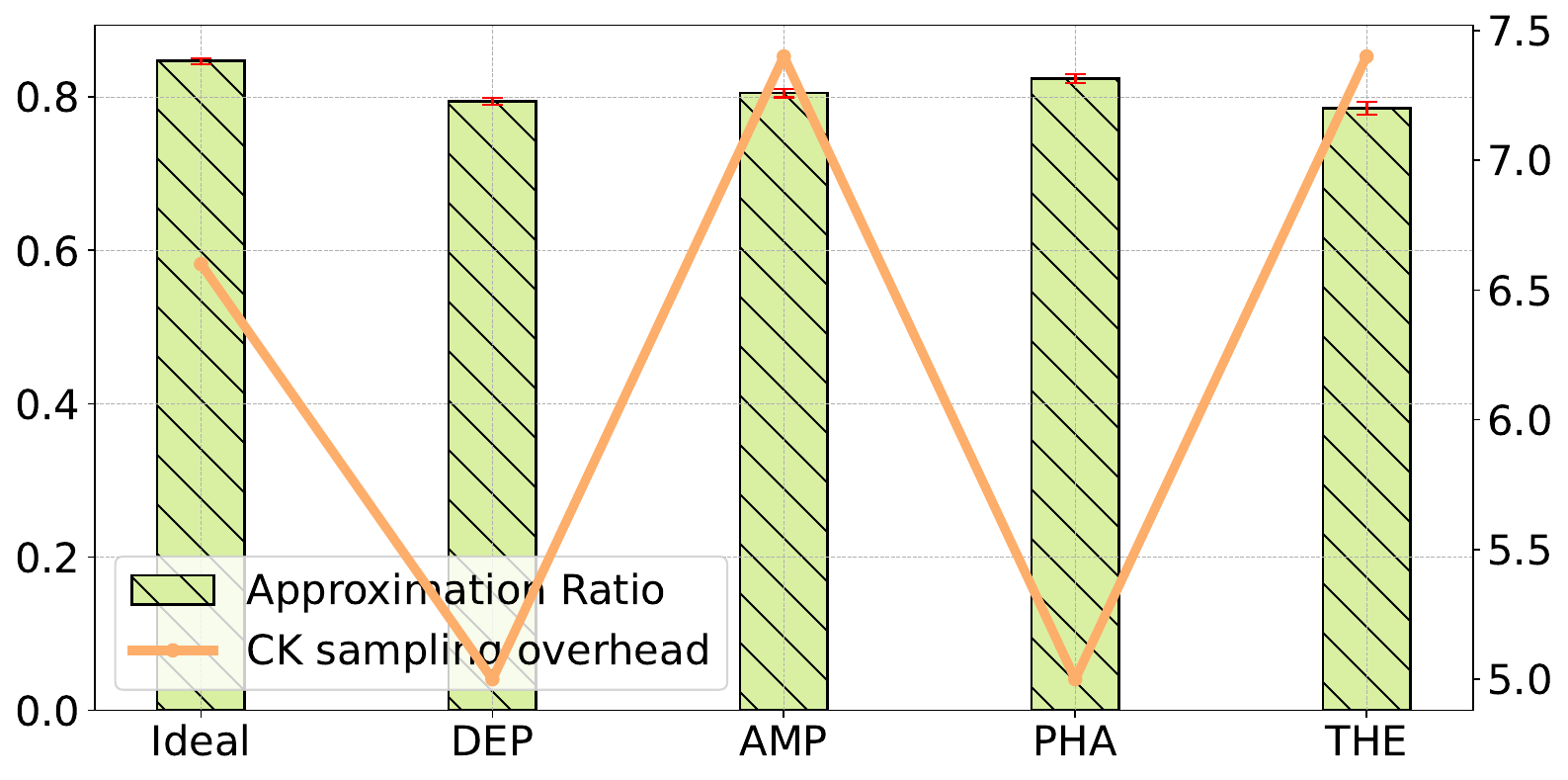}
    \caption{Performance of the proposed method under noiseless simulator and 4 different quantum noise simulators.}
     \label{fig:noise-performance}
     \Description{}
\end{figure}

To evaluate robustness in realistic NISQ settings, we test \CKVQA\ under four standard noise models---depolarizing (DEP), amplitude damping (AMP), phase damping (PHA), and thermal relaxation (THE)---each with noise probability 0.01. QAOA is applied to Max-Cut on 8-node 3-regular graphs, and performance is reported as the cut-value ratio relative to the optimum. All settings are repeated 10 times to reduce stochastic variance.
As shown in Fig.~\ref{fig:noise-performance}, \CKVQA\ delivers approximation ratios close to those of the noiseless simulator across all noise types, indicating that its circuit-cutting and optimization pipeline is resilient to moderate device noise. These results suggest that \CKVQA\ remains effective on current NISQ hardware without requiring noise-tailored modifications.

%% file: chapter/7-conclusion.tex


This paper presents \CKVQA, a co-design framework that integrates circuit cutting with cutting-aware QAS to automatically generate ansatz structures for VQAs. The QAS module identifies ansätze that satisfy user-defined sampling-overhead constraints while maintaining algorithmic accuracy, and subcircuit-level parameter-localized optimization minimizes execution time without degrading solution quality. Experiments demonstrate that \CKVQA\ substantially reduces sampling overhead, accelerates VQA execution, and remains robust under representative noise models. 
The results highlight that scalable VQA execution fundamentally requires architecture-aware ansatz design, suggesting that future multi-QPU systems should co-optimize circuit structures with partitioning strategies. 
Future work includes combining \CKVQA\ with advanced compilation and scheduling techniques and leveraging circuit-cutting information to prune the QAS search space, enabling faster and more scalable ansatz discovery independent of hardware constraints.